\documentclass{aa}
\usepackage{txfonts,textcomp}

\usepackage{natbib}
\bibpunct{(}{)}{;}{a}{}{,}

\usepackage{graphicx}
\usepackage[colorlinks=true, allcolors=blue]{hyperref}
\usepackage{adjustbox}

\begin{document}

    \title{Joint machine learning and analytic track reconstruction for X-ray polarimetry with gas pixel detectors}

    \author{N. Cibrario\inst{1,2}
        \and
        M. Negro\inst{3,4,5}
        \and
        N. Moriakov\inst{6}
        \and
        R. Bonino\inst{1,2}
        \and
        L. Baldini\inst{7,8}
        \and
        N. Di Lalla\inst{9}
        \and
        L. Latronico\inst{1}
        \and
        S. Maldera\inst{1}
        \and
        A. Manfreda\inst{7,10}
        \and
        N. Omodei\inst{9}
        \and
        C. Sgr\'o\inst{7}
        \and
        S. Tugliani\inst{1,2}  
        }

    \institute{Istituto Nazionale di Fisica Nucleare, Sezione di Torino, Via Pietro Giuria 1, 10125 Torino, Italy
        \and
        Dipartimento di Fisica, Università degli Studi di Torino, Via Pietro Giuria 1, 10125 Torino, Italy
        \and
        University of Maryland, Baltimore County, Baltimore, MD 21250, USA
        \and
        NASA Goddard Space Flight Center, Greenbelt, MD 20771, USA
        \and
        Center for Research and Exploration in Space Science and Technology, NASA/GSFC, Greenbelt, MD 20771, USA
        \and
        Department of Radiation Oncology, Netherlands Cancer Institute, the Netherlands
        \and
        Istituto Nazionale di Fisica Nucleare, Sezione di Pisa, Largo B. Pontecorvo 3, 56127 Pisa, Italy
        \and
        Dipartimento di Fisica, Università di Pisa, Largo B. Pontecorvo 3, 56127 Pisa, Italy
        \and
        Department of Physics and Kavli Institute for Particle Astrophysics and Cosmology, Stanford University, Stanford, California 94305, USA
        \and
        Istituto Nazionale di Fisica Nucleare, Sezione di Napoli, Strada Comunale Cinthia, 80126 Napoli, Italy
        }

\abstract{We present our study on the reconstruction of photoelectron tracks in gas pixel detectors used for astrophysical X-ray polarimetry. Our work aims to maximize the performance of convolutional neural networks (CNNs) to predict the impact point of incoming X-rays from the image of the photoelectron track.
A very high precision in the reconstruction of the impact point position is achieved thanks to the introduction of an artificial sharpening process of the images.
We find that providing the CNN-predicted impact point as input to the state-of-the-art analytic analysis improves the modulation factor ($\sim 1 \%$ at 3 keV and $\sim 6 \%$ at 6 keV) and naturally mitigates a subtle effect appearing in polarization measurements of bright extended sources known as "polarization leakage".}


\maketitle

\section{Introduction}

Linear X-ray polarization is of interests to a wide range of fields in physics: from crystal dynamical diffraction \citep{diffr}, to polarization radiography to supplement mammographic images \citep{mammography}, and to high-energy astrophysics. The latter application, in particular, is the focus of this work, but the outcome of this study can be easily extended to other applications based on the same detection technique. 

The first astrophysical X-ray polarization measurements date back to the 1970s with the observation of the Crab Nebula \citep{CrabNebulapolarization_novick, CrabNebulaPolarization_Weisskopf}. Such polarization measurements were based on Bragg reflection \citep{BraggPolarimetry} at 45 degrees of the X-ray on a crystal, exploiting the Bragg diffraction dependence on the radiation polarization (only the X-rays polarized perpendicularly to the plane of incidence are reflected).   

X-ray polarimetry can be done significantly more efficiently exploiting the high dependence of the photoelectric effect on the polarization of the incident radiation. This technique was proposed in early 2000 \citep{articleCosta}, and reached full maturity with the gas pixel detector (GPD) \citep{BellazziniGPD}, now acquiring data on board the \textit{Imaging X-ray Polarimetry Explorer} (\textit{IXPE}) which was launched by NASA in 2021 \citep{IXPE_calibration}, and it will be installed on the future Chinese mission, \textit{} \textit{enhanced X-ray Timing and Polarimetry }(eXTP) \citep{eXTP}. Such an instrument combines good imaging capabilities and unprecedented polarization sensitivity and has already opened a new path for the future of astrophysics. The typical energy range of these focusing X-ray polarimeters is between 1 and 10 keV, with a highly variable effective area within this range depending on the instrument focusing optics and detectors\footnote{Specifically, IXPE is optimized and calibrated to work between 2 and 8 keV.}. Given the power-law nature of the spectra of virtually all astrophysical X-ray sources, it is crucial to have the best polarization sensitivities in the lower end of the energy bandwidth. For the reasons we illustrate in the next section, the X-ray polarization direction of the lower energy X-rays is also the most difficult to measure with the GPD. Indeed, at low energies all the present reconstruction methods, both the state-of-the-art analytic one developed by the IXPE collaboration and the recently developed machine learning (ML) techniques, show the biggest limitations. Moreover, all these reconstruction strategies suffer from a systematic effect called "polarization leakage" \citep{leakage}, which we briefly discuss in Sec.~\ref{sec:3}. 

 In this paper we propose a hybrid analytic-ML approach, in which we exploit a ML algorithm based on a convolutional neural network (CNN) to improve the performance of the state-of-the-art analytic algorithm, and to mitigate the polarization leakage effect. The working principle of the GPD and the data set used for the analysis are discussed in Sec.~\ref{sec:2}. In Sec.~\ref{sec:3} the overall features of the reconstruction methods are depicted, with a focus on the relevance of the impact point parameter. In Sec.~\ref{sec:4} we describe the structure of the adopted CNN and illustrate its training, optimization processes, and the results regarding the reconstruction of the impact point location. In Sec.~\ref{sec:5} we present the polarization results we obtain with our hybrid algorithm. Sec.~\ref{sec:6} summarizes the results and provides considerations for future applications and developments. \\

\section{Instrument and data set}
\label{sec:2}

\begin{figure}[htb]
    \centering
    \includegraphics[width=7cm, height=7cm]{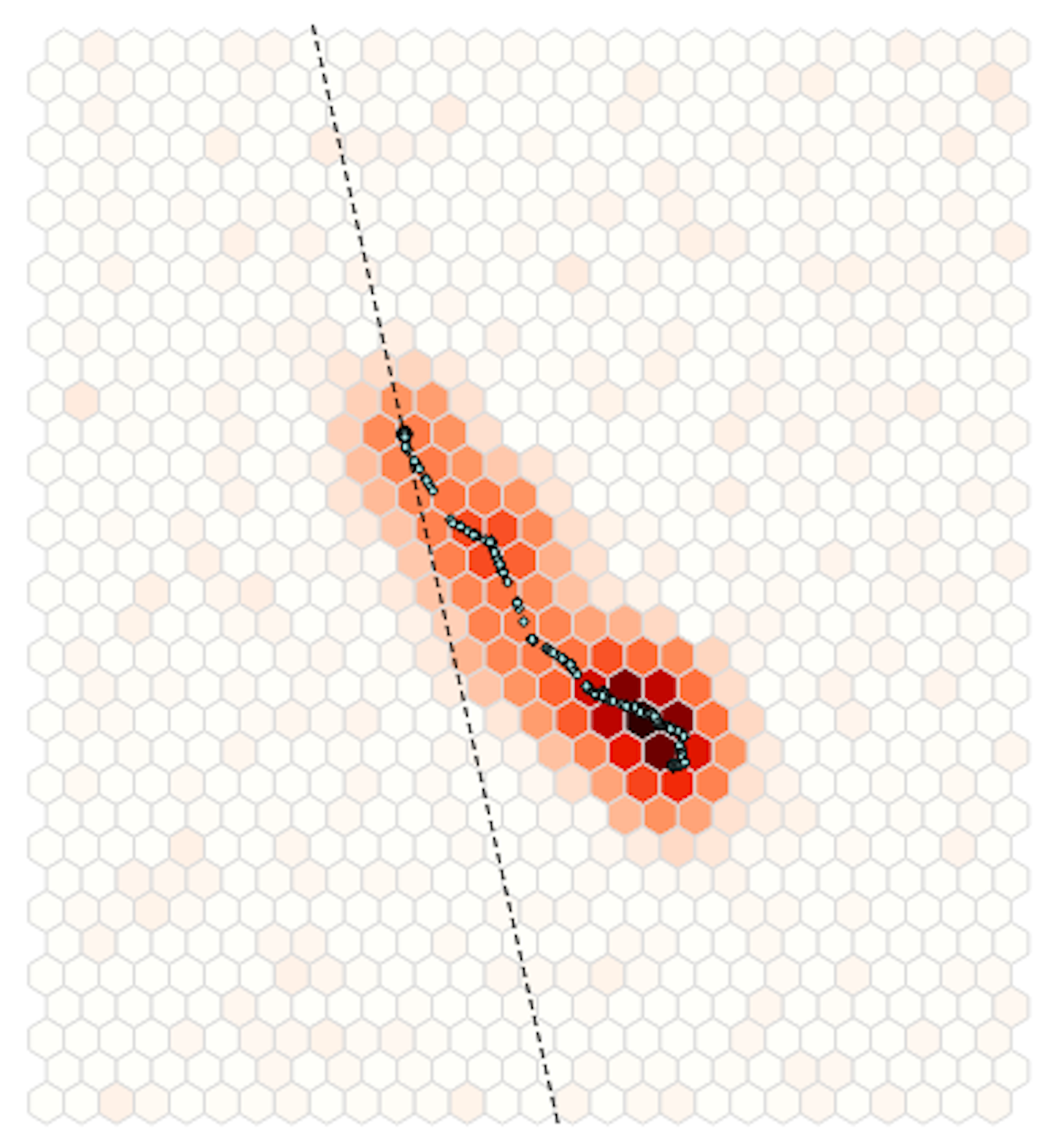}
    \includegraphics[width=7cm, height=7cm]{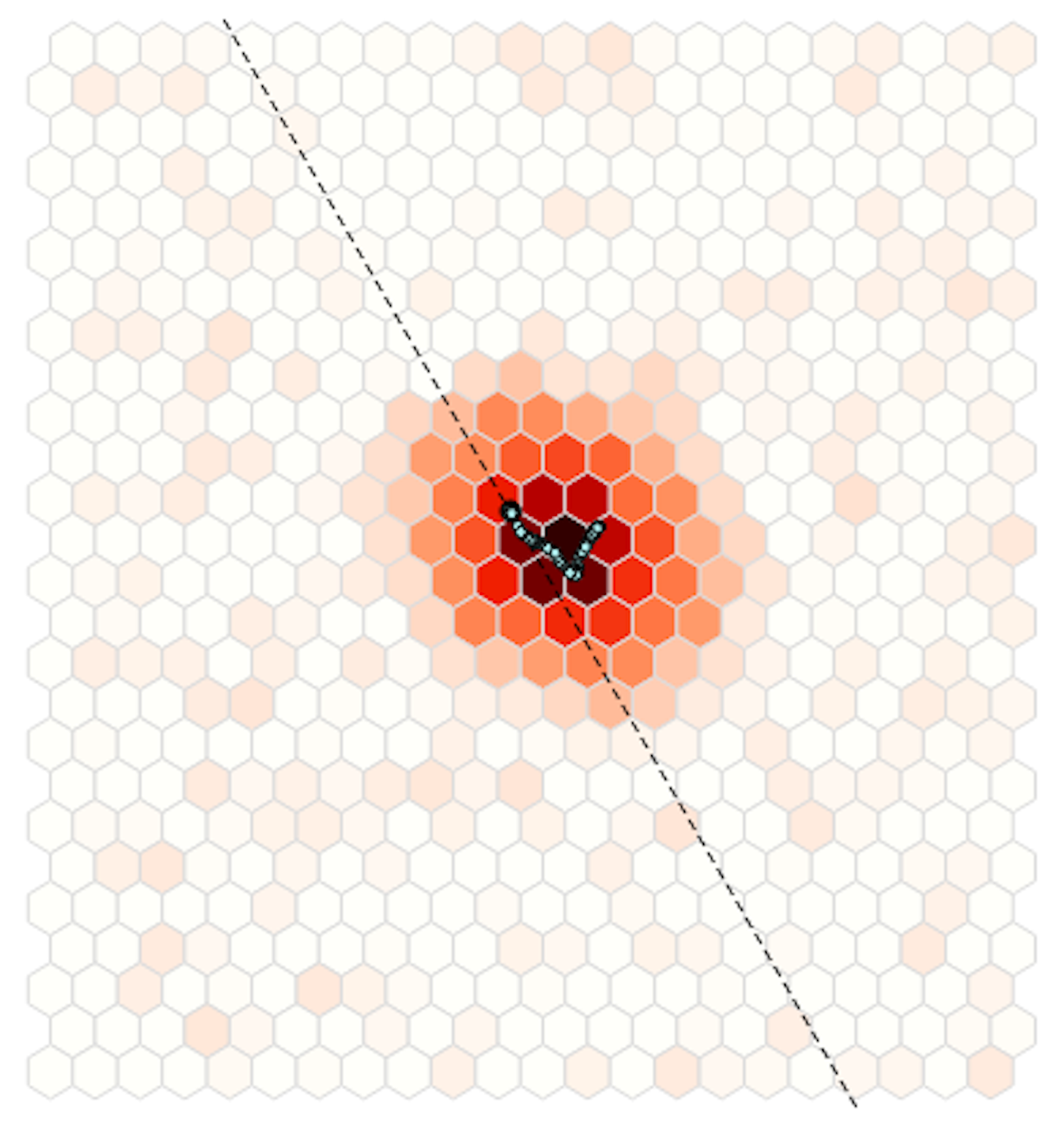}
    \caption{Example of a photoelectron track generated by a 7 keV photon (top panel) and a 3.5 keV photon (bottom panel). The dashed lines and the black points represent the simulated PE emission direction and X-ray impact point, respectively. The simulated photoelectron path is reported as well. The color intensity of the pixels is proportional to the energy deposited in the gas by the photoelectron.  
    }
    \label{fig:tracks}
\end{figure}

The complete description of how the GPD functions can be found in~\cite{BellazziniGPD}. Here we only summarize the general concept behind the use of GPDs to measure the X-ray polarization. As mentioned in the introduction, the instrument functioning is based on the photo-electric effect, in which an X-ray is absorbed in the gas gap of the GPD\footnote{The gas mixture used for IXPE is Dimethyl Ether ($\rm (CH_3)_2O$).}, and a photoelectron (PE) is ejected in the direction ($\theta, \phi$), namely the emission direction, which preferably lies on the oscillation plane of the electric field of the incoming X-ray. We note that $\theta$ is the angle between the incident X-ray direction and the PE emission direction, while $\phi$ is the azimuthal PE emission direction. The PE interacts with the gas atoms through ionizing collisions, losing energy at each collision\footnote{The energy loss $\frac{\partial E}{\partial x}$ is inversely proportional to the kinetic energy of the electron: $\frac{\partial E}{\partial x} \propto \frac{1}{\beta^2} \propto \frac{1}{E_{{\rm }kin}}$, where $\beta$ is the velocity of the electron in units of c and $E_{{\rm }kin}$ is its kinetic energy.}. Such interactions generate a pattern of ion-electron pairs called "track" that marks the path followed by the PE before losing all its energy and being reabsorbed in the gas. The primary e$^-$ charges are  amplified and collected on a plane of hexagonal pixels in a honeycomb configuration. A track, therefore, is a pixellated image containing useful information about the PE, and, as a consequence, about the X-ray that generated that same PE. Two examples of a PE track image for two different X-ray energies are reported in Fig.~\ref{fig:tracks}.

Our data set consists of PE tracks generated through a Monte Carlo (MC) simulation software, named "ixpesim"~\citep{niccolo}, which relies on GEANT4 and uses a slightly customized version of the "Livermore Polarized" physics list. The data set consists of three different categories of simulations.

Firstly, we generated two million events from an unpolarized beam, with a flat energy spectrum in the range 1.0-9.0 keV. This energy range includes the energies of the highest sensitivity (mostly due to the higher effective area) of current and future experiments adopting this X-ray polarization technique, such as IXPE and eXTP. This portion of the data set was used to train and validate our CNN. The simulated gas pressure was set at 720 mbar for all the events we generated. Secondly, we generated test samples by simulating 100,000 events for each set at fixed energies: two sets (100\% polarized and unpolarized) for 13 different mono-chromatic beams (between 2 keV and 8 keV with a 0.5 keV step). These events were used to evaluate and compare the performance of the algorithms for different energy values of the incident X-rays. Finally, we generated three sets of 500,000 events each, simulating three different unpolarized point sources with typical spectral shapes of astrophysical sources. Specifically, we simulated two power-law spectra (with -0.7 and -1.7 spectral indices) and one blackbody spectrum. This data set was used to evaluate and study the polarization leakage effect. For each MC track, we know the true polarization angle, the true X-ray energy and impact point, and the true PE emission angle.\\

\section{Reconstruction methods}
\label{sec:3}
The event reconstruction consists of the estimation of the properties of interest of the incoming X-ray photon through the PE track. In particular, the total collected charge is proportional to the energy of the X-ray, the starting point of the track gives the impact point (IP) of the photon, and the azimuthal angle $\phi$ of the PE emission direction (before it gets deviated by multiple interactions in the gas) carries the memory of the X-ray polarization direction. While the latter is the track parameter that directly provides the actual information about the polarization properties of the incident X-rays\footnote{The polarization degree and angle were statistically estimated on the basis of the PE azimuthal distribution.}, the photon impact point represents a key feature in the reconstruction process, both to avoid biases and to improve the general performance of the algorithms.  

As shown in Fig.~\ref{fig:tracks}, the track morphology strongly depends on the energy of the absorbed X-ray. For low energy X-rays, the PE path is generally a few pixels long, and the track is essentially round. The lack of elongation, as well as the absence of skewness in the spatial distribution of the charge make the reconstruction of the properties of the tracks generated by low energy X-rays more challenging.

\subsection{Analytic reconstruction with moment analysis}
\label{sec:ma}
The state-of-the-art algorithm currently used by the IXPE collaboration~\citep{AnalyticRecon} is based on an analysis of the momenta of the track image, and hence called moment analysis. In short, this algorithm relies on the morphological properties, especially the elongation, of the track as well as on the deposited energy in each pixel to estimate both the impact point of the X-ray and the PE emission direction. 

\begin{figure}[ht]
    \centering
    \resizebox{\hsize}{!}{\includegraphics[width=8cm, height=8cm]{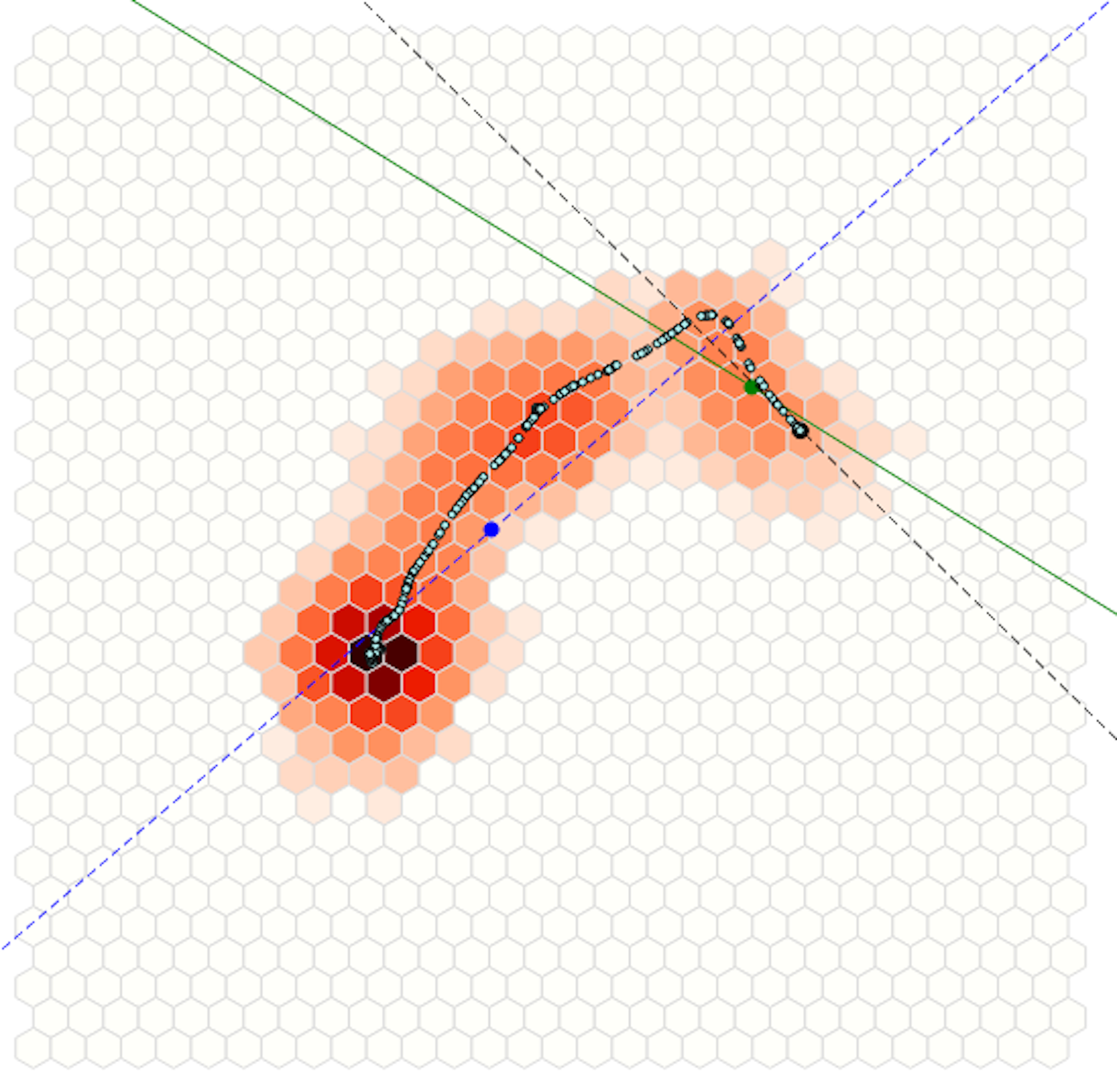}}
    \caption{Example of a 7.5 keV PE track with all the parameters reconstructed with moment analysis. The blue point and the dashed blue line are the barycenter and the direction of maximum elongation of the track, respectively. The green point and the green line are the reconstructed impact point and the reconstructed emission direction, respectively. The black point and the dashed black line are the true (MC) impact point and emission direction, respectively.}
    \label{fig:rec_example}
\end{figure}

The details of the process are reported in~\cite{AnalyticRecon} and summarized in Appendix \ref{app:b}. Here we report only the final step of the analysis. Once the impact point parameter was reconstructed, the second moment of the charge distribution $\rm M'_2(\phi)$ with respect to the location of the impact point was calculated as follows:
\begin{equation}
    M_2'(\phi)=\frac{\sum_{\rm{i}} w_{\rm{i}} [(x_{\rm{i}} - x_{\rm{IP}}) \cos(\phi) + (y_{\rm{i}} - y_{\rm{IP}}) \sin(\phi)]^2}{\sum_{\rm{i}} w_{\rm{i}}}
    \label{eq:mom_analysis}
,\end{equation}
\noindent
where ($x_i$, $y_i$) is the position of each $i_{th}$ pixel, and $w_i$ is the weight assigned based on the skewness of the track. The emission angle was obtained by evaluating the angle $\phi$ with respect to the $x$ axis which maximized $M_2'(\phi)$. In Fig.~\ref{fig:rec_example} a track image along with the parameters reconstructed with moment analysis is shown.

\subsubsection{Impact point and modulation factor}
\label{sec::key1}
The impact point of the incident photon is not directly linked to its polarization properties, but it plays an important role in the emission angle reconstruction with moment analysis. An incorrect estimation of the IP inevitably affects the subsequent reconstruction of the emission angle, as it is involved in Eq.~\ref{eq:mom_analysis}. 

In order to compare the performance of different algorithms, we used a figure of merit, the "modulation factor". Denoted as $\mu$, it represents the response of a polarimeter to a 100\% polarized source as the normalized half-counting rate difference. It ranges between zero (no sensitivity) and one (maximum sensitivity). Its calculation through Stokes parameters is described in ~\cite{Stokes}. The sensitivity of a polarimeter is affected both by the instrument limitations, and by the performance and the efficiency of the reconstruction algorithm. In this work we focus on the effects linked to the reconstruction algorithm.

\begin{figure}[ht]
    \centering
    \hspace*{-0.4cm}
    \resizebox{\hsize}{!}{\includegraphics[width=10cm, height=7cm]{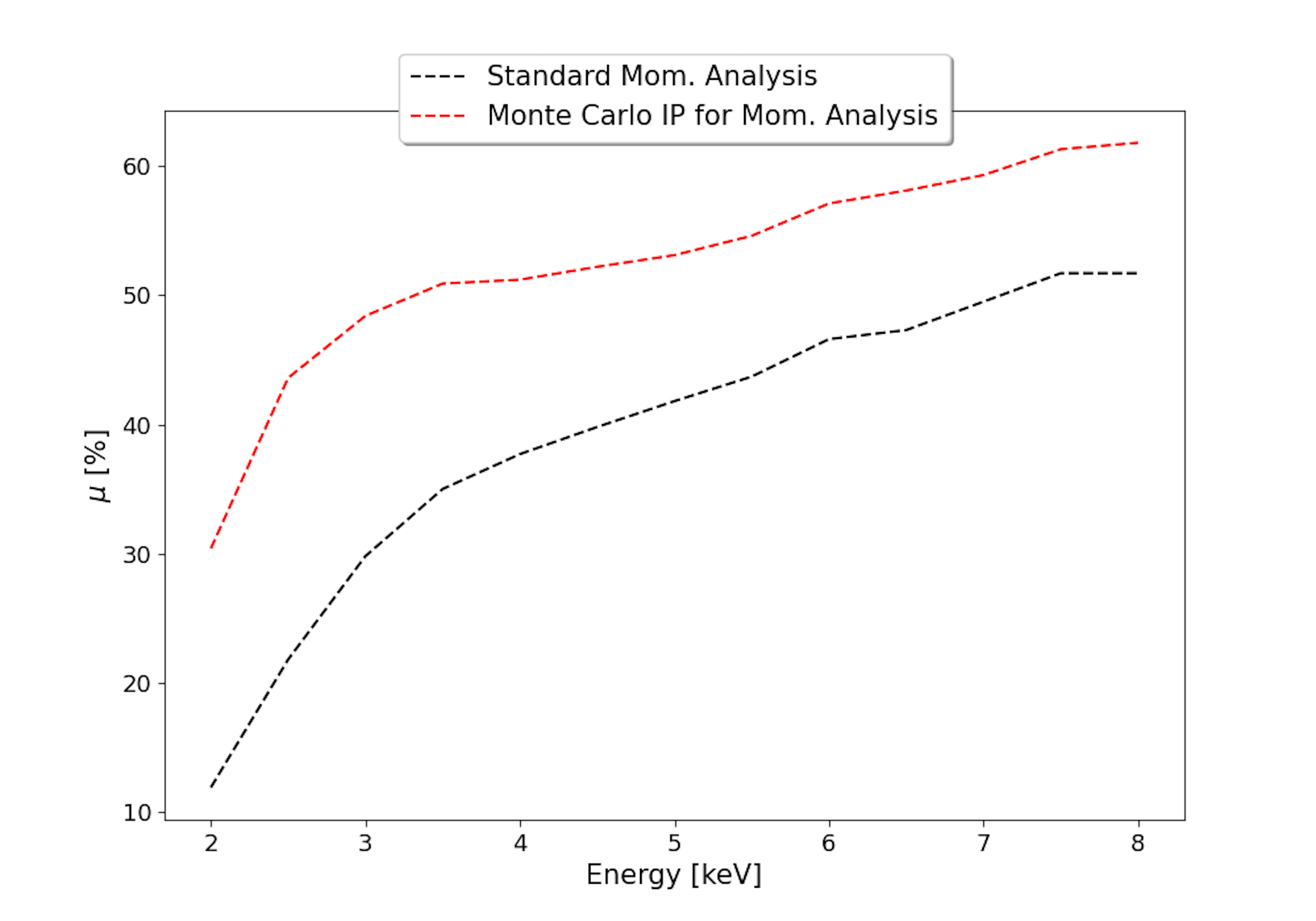}}
    \caption{Modulation factor as a function of energy for the standard moment analysis (black) and for the same analysis but substituting the predicted IP with the true one (red).}
    \label{fig:ip_mom}
\end{figure}

In Fig.~\ref{fig:ip_mom} the modulation factor as a function of the energy is reported in black for the standard moment analysis. The worsening of $\mu$ with decreasing energy is due to the difficulties in reconstructing the correct emission angles of low energy photoelectron tracks. In the same figure, the red line represents the modulation factor obtained by substituting the predicted impact point with the MC (true) one in Eq.~\ref{eq:mom_analysis}. This substitution results in a significant improvement in the performance of the algorithm. Clearly this procedure cannot be applied to experimental data, as we do not know the true IP location, but it highlights how improving the precision in the prediction of the IP position could lead to improved precision in the determination of the emission angle.\\

\subsubsection{Impact point and polarization leakage}
\label{sec::key2}

Recently, a systematic effect denominated polarization leakage has been found in IXPE measurements and discussed in \cite{leakage}. Due to the poor reconstruction of the correct impact point from PE tracks, some astrophysical sources exhibit an induced polarization pattern associated with intensity edges and gradients. 

\begin{figure}[ht]
    \centering
    \hspace*{-0.4cm}
    \includegraphics[width=7cm, height=6cm]{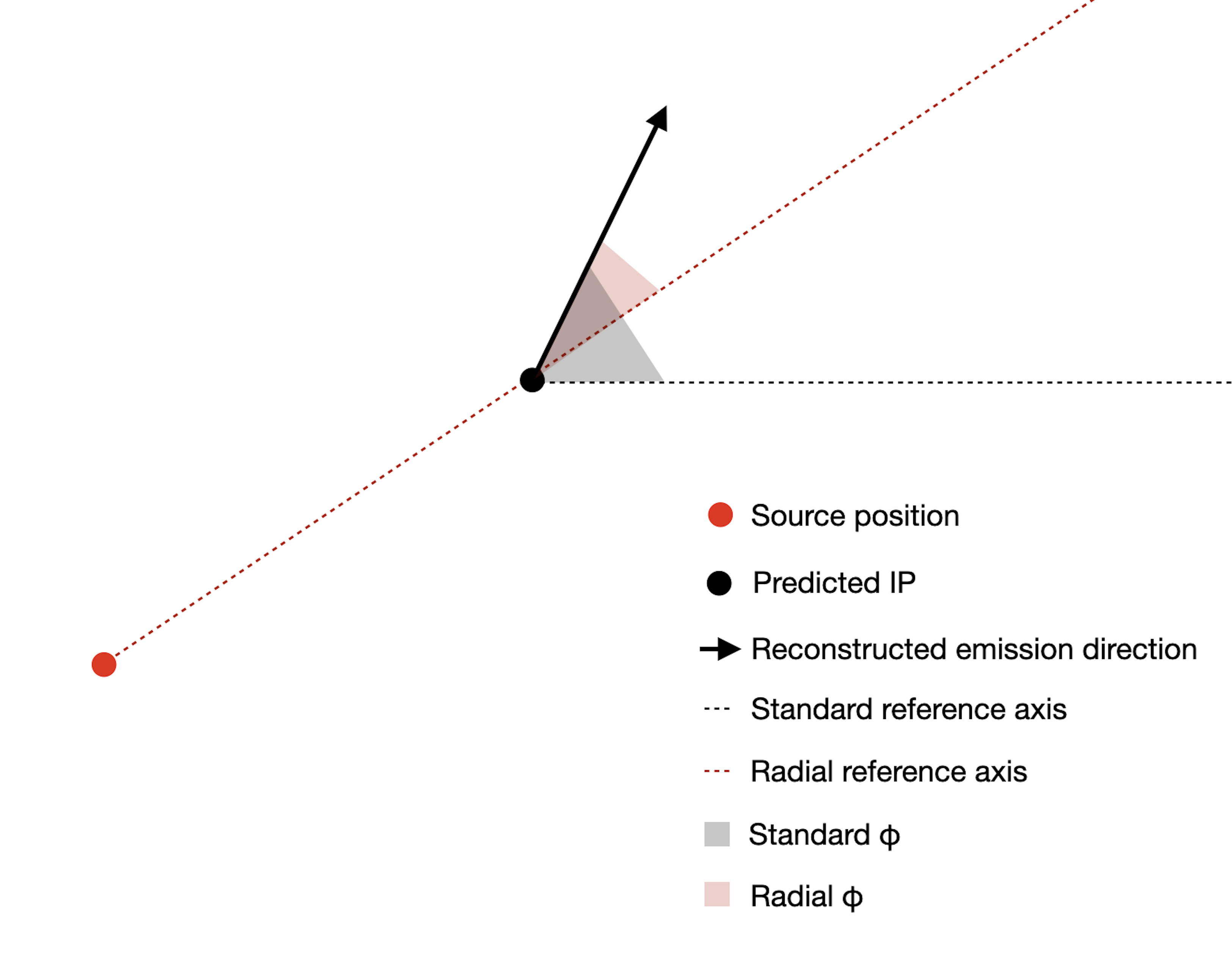}
    \caption{Representation of the radial alignment for a single event. The red and black dots are the source and the predicted IP positions, respectively. The black arrow is the predicted emission direction. The dashed red and black lines are the radial and standard reference axes, respectively, while the red and black angles are the radial and standard predicted emission angles, respectively.}
    \label{fig:align}
\end{figure}

This effect can be particularly appreciated when analysing unpolarized point sources. For these sources polarization leakage can cause an induced radial polarization. For this study we thus considered the three simulated unpolarized point sources described in Sec. \ref{sec:2}. In order to observe a potential induced radial polarization -- since the pattern is, by definition, radially symmetric around the center of the point source -- we measured the PE emission angle with respect to the reference axis defined by the radial direction of the impact point. A schematic representation of the radial alignment is reported in Fig. \ref{fig:align}.

Firstly, for each event, we calculated the radial direction as follows:

\begin{equation}
    \phi_{0} = \tan^{-1}\left(\frac{x_{\rm{rec}}-x_{\rm{s}}}{y_{\rm{rec}}-y_{\rm{s}}}\right)
,\end{equation}
\noindent
where ($x_{\rm{rec}}-x_{\rm{s}}$) and ($y_{\rm{rec}}-y_{\rm{s}}$) are the horizontal and vertical distances of the reconstructed IP from the simulated source position. In our case, $(x_{\rm{s}},y_{\rm{s}})=(0,0)$. We then aligned the Stokes parameters to the radial direction,

\begin{equation}
    Q_{\rm{rad}} = \frac{1}{2} Q_{\rm{rec}} Q_0 + \frac{1}{2} U_{\rm{rec}} U_0 
\end{equation}
\begin{equation}
    U_{\rm{rad}} = \frac{1}{2} U_{\rm{rec}} Q_0 - \frac{1}{2} Q_{\rm{rec}} U_0 
,\end{equation}
\noindent
where $Q_{\rm{rec(0)}}=2\cos(2\phi_{\rm{rec(0)}})$, $U_{\rm{rec(0)}}=2\sin(2\phi_{\rm{rec(0)}})$,  and $\phi_{\rm{rec}}$ denotes the reconstructed emission angle direction. From the Stokes parameters, we calculated the polarization degree and angle as described in \cite{Stokes}.
The residual radial modulation obtained with the standard moment analysis for our sources is reported in Table~\ref{tab:0}.

\begin{table}[ht]
\centering
\caption{Summary of the residual radial modulation calculated with standard moment analysis for the unpolarized point sources.}
\begin{adjustbox}{width=0.47\textwidth}
\begin{tabular}{cc}     
\hline
\hline
\noalign{\smallskip}
\textbf{Spectral Model} & \textbf{Residual Radial Modulation} [\%] \\
\hline
\noalign{\smallskip}
Power Law, index -1.7 (PL1) & $6.27 \pm 0.31$ \\
\noalign{\smallskip}
Power Law, index -0.7 (PL2) & $6.48 \pm 0.26$ \\
\noalign{\smallskip}
Blackbody, kT 3keV (BB) & $6.58 \pm 0.23$ \\
\hline
\end{tabular}
\end{adjustbox}
\label{tab:0}
\end{table}

\begin{figure}[ht]
    \centering
    \hspace*{-1.2cm}
    \includegraphics[width=11cm, height=5cm]{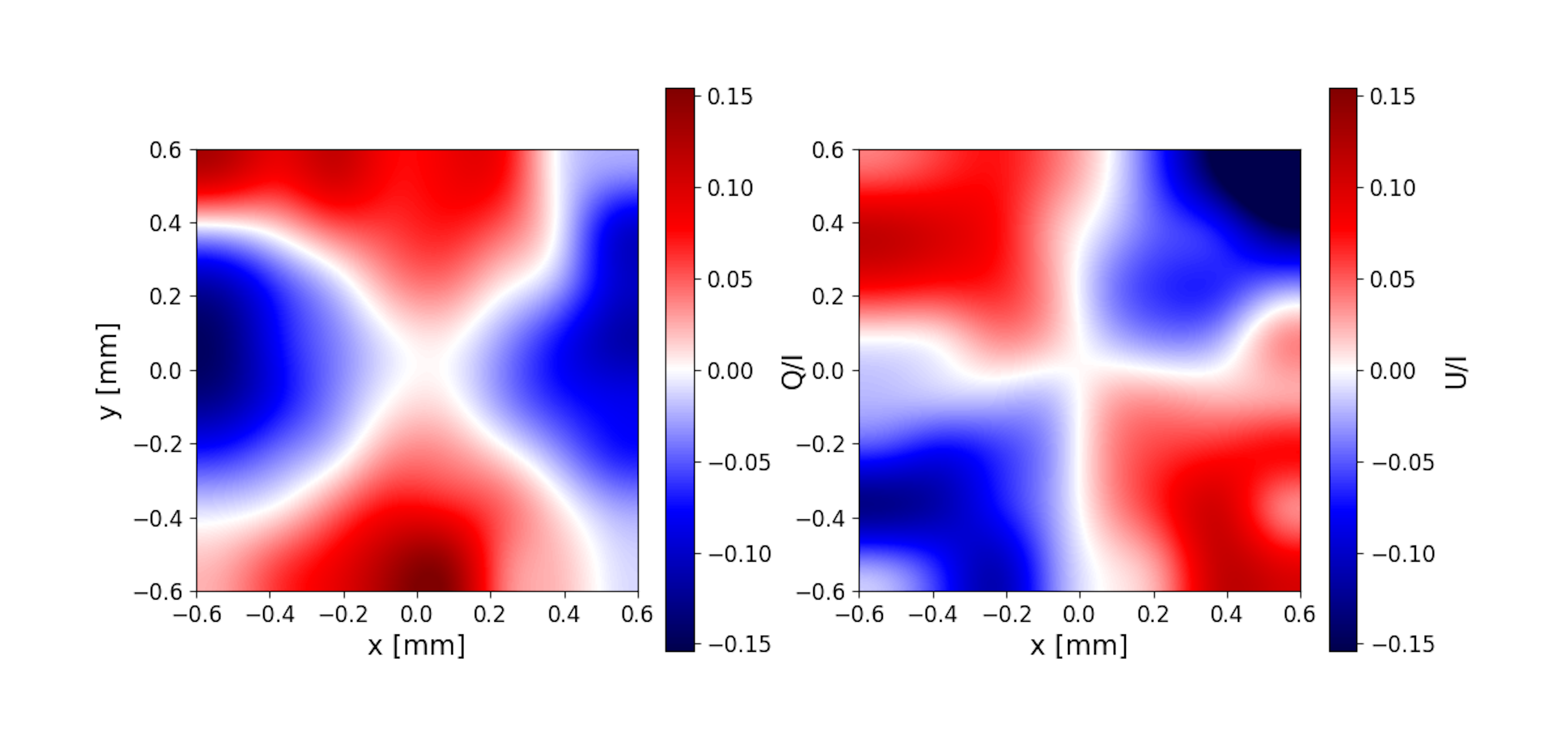}
    \caption{Binned and interpolated calculation of Q/I and U/I, with Q, U, and I being the Stokes parameters, for the blackbody spectrum source (BB). The source is located at the center of the GPD, $(x_s,y_s) = (0,0)$.}
    \label{fig:QU}
\end{figure}

From Table~\ref{tab:0} we notice that the moment analysis finds residual radial modulations up to $6.58 \pm 0.23 \%$, even if the simulated sources are unpolarized. The same effect can be noticed by binning the area surrounding the source and evaluating the quantities Q/I and U/I, with Q, U, and I being the Stokes parameters. In Fig.~\ref{fig:QU} we report the results for the blackbody spectrum source. We would again expect values compatible with zero,  while we can notice a clear pattern that indicates a residual radial polarization.

\subsection{CNN-based algorithms}
Several groups have already shown the potential of ML to reconstruct the PE track and determine the incident photon properties. For example,~\cite{TakaoCNNclassification} trained a CNN to learn how to classify the PE tracks into polarization angle bins, and the same NN also predicts the impact point position on the track. ~\cite{OurCNN} tested the use of CNNs through regression instead of classification, again to infer the impact point and the  PE initial emission direction. \cite{StanfordCNN} and \cite{ROMANI2021}  also used CNNs to predict the energy, impact point, and polarization direction, evaluating prediction uncertainties for each event thorough a deep ensemble technique \citep{DeepEnsembles}. In all of these works, the CNN architecture and the preprocessing phase of the images were focused on the reconstruction of the emission angle, with the IP prediction being a "secondary" product of the reconstruction. 

Our work is intended to build and optimize a CNN to specifically reconstruct the IP location. By substituting the IP predicted by moment analysis with the one predicted by the CNN in Eq.~\ref{eq:mom_analysis}, we aim both to improve the performance of the algorithm and to mitigate some systematic effects, such as the polarization leakage effect. \\

\section{CNN for the impact point reconstruction}
\label{sec:4}

\subsection{The CNN architecture}

We built a CNN based on the DenseNet-121~\citep{DenseNets} architecture, which we modified to incorporate hexagonal convolutions at minimum performance cost. We summarize the key features of our CNN below.

We started with implementing hexagonal convolution layers as a C++ extension for PyTorch, since a reference Python-only implementation of hexagonal convolutions in~\cite{hexagdly_paper} is substantially slower compared to the standard 2D Cartesian convolutions. Incorporating hexagonal convolutions allows the network to correctly capture the track structure and the spatial dependencies between pixels on the hexagonal pixel grid, which is particularly important in the initial layers of the CNN. Therefore, we employed hexagonal convolutions in the first convolution block, which consists of a stack of three hexagonal convolution layers with 64 filters in each layer, stride 1 and kernel radius 1. Each convolution layer here is followed by a batch normalization layer and rectified linear unit (ReLU) activation function.

For performance reasons, after the initial hexagonal convolution block, a transition was performed from the hexagonal grid to the Cartesian grid by applying a hexagonal convolution layer on a Cartesian subgrid of the hexagonal grid with stride 2. This "transition convolution" has kernel radius 1 and 64 filters as well, where the combination of stride 2 and kernel radius 1 implies that this transition convolution gathers image features from the entirety of the hexagonal grid. The transition convolution is followed by a batch normalization layer and ReLU activation. Subsequently, we switched to the standard DenseBlocks of the DenseNet-121 using Cartesian 2D convolutions only. A fully connected layer for impact point regression was applied to the resulting final CNN feature map with dimensions $6 \times 6$. Throughout the network, dropout layers with a dropout probability of $10 \%$ were used inside the DenseBlocks. 

The network was built to predict the impact point position $(x_{\rm{true}},y_{\rm{true}})$. The information is given in the number of decimal pixels ($0 < x_{\rm{true}}(y_{\rm{true}}) < 72$), and we used the following loss function:

\begin{equation}
    \label{eq:lossfunc}
     L(x_{\rm{true}},y_{\rm{true}} | x_{\rm{pred}},y_{\rm{pred}}) = |(x_{\rm{true}},y_{\rm{true}})-(x_{\rm{pred}},y_{\rm{pred}})|
.\end{equation}

We used the Adam optimizer \citep{adam}, a commonly employed type
of stochastic gradient descent, with a decaying learning rate starting from $10^{-4}$. A form of online hard example mining (OHEM) \citep{ohem} was employed during training to improve the general performance of the network, where in each batch only $50\%$ of samples with the worst performance was taken into account during gradient computation.

\subsection{Preprocessing, optimization, and training of the CNN}
As described in Sec.~\ref{sec:2}, our data set consists of PE tracks generated through MC simulations. A preprocessing of the track images was needed in order to prepare the sample for the CNN. 

The characteristic detector noise was realistically taken into account in the simulations and generated a background of nonzero value pixels in the frame of the track images. To suppress such a noise, we set all pixels with values below a threshold of 20 analog to digital converter (ADC) counts  to zero (which corresponds to $\sim 45$ electrons)~\citep{Coordinate}. Additionally, pixels with values above the threshold but disconnected from the main pixel cluster of the track were also set to zero. This process facilitated and accelerated the CNN training by removing useless information. However, we also tested a CNN with noisy images and the performance was not significantly worse: the CNN on its own learns how to ignore the information carried by the random pixel noise. The pixel values were then rescaled to between the range of zero and one. 

\begin{figure}[ht]
    \centering
    \includegraphics[width=7cm, height=6cm]{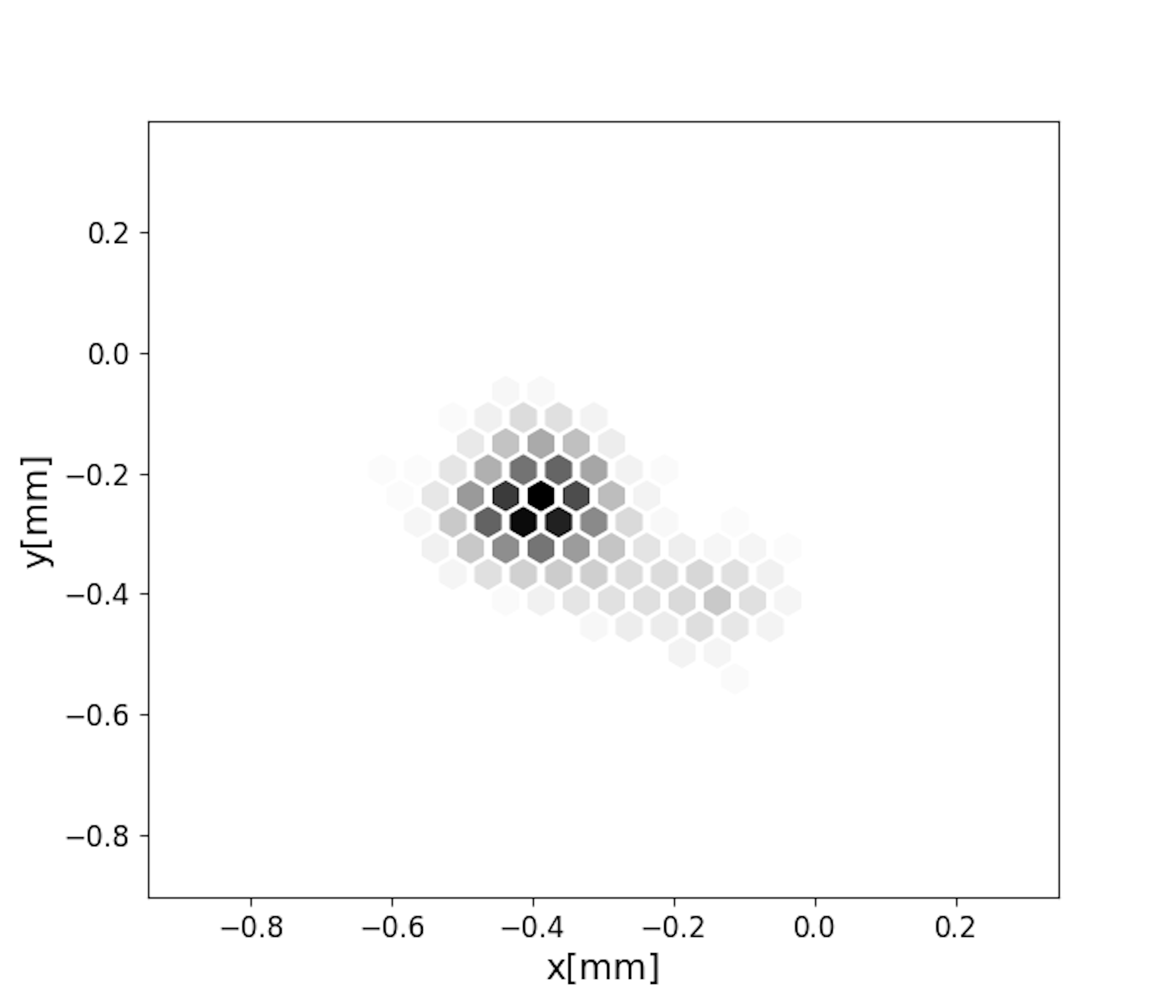}
    \includegraphics[width=7cm, height=6cm]{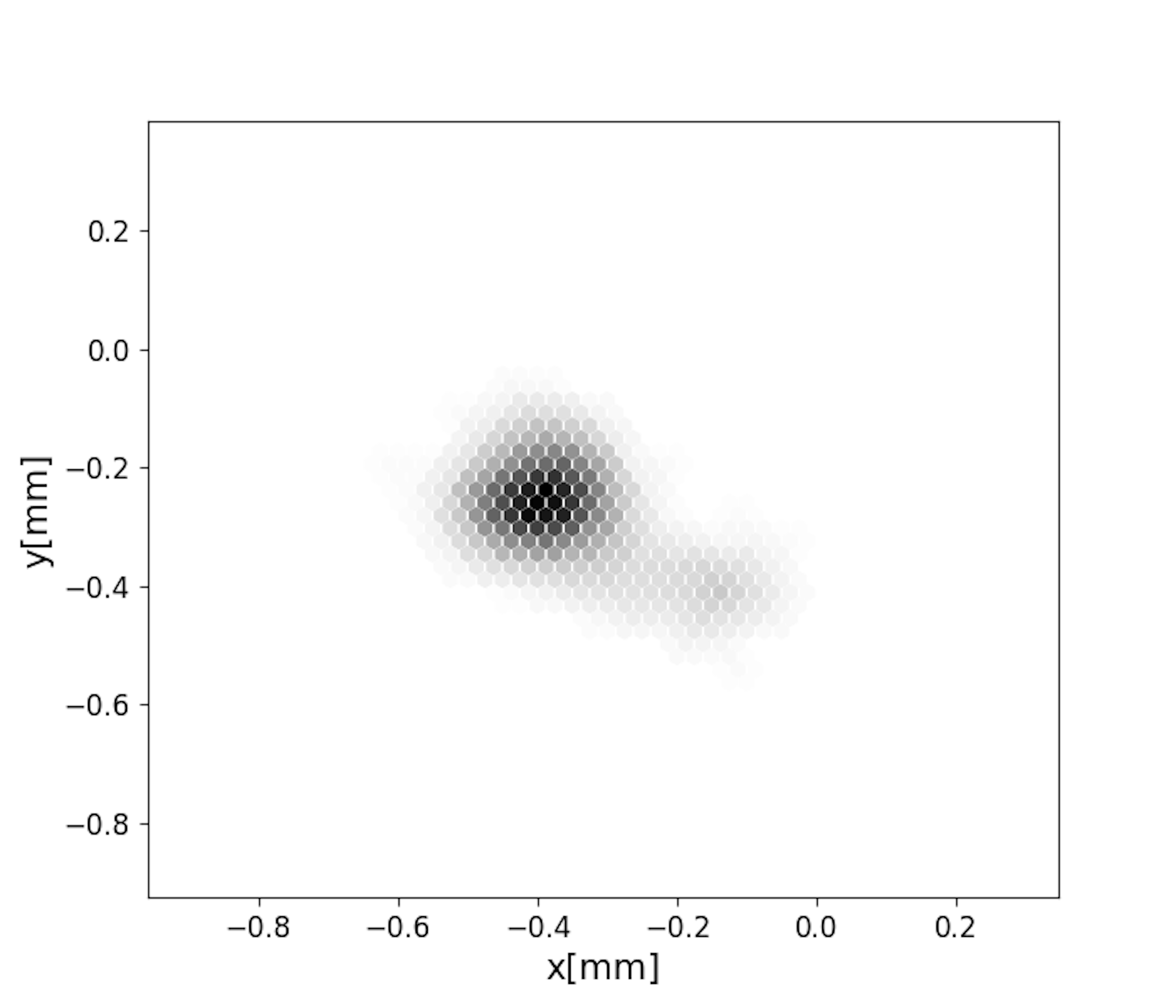}
    \caption{Example of a PE track before (top panel) and after (bottom panel) the sharpening process.  
    }
    \label{fig:comp}
\end{figure}

In order to improve the precision of the network in identifying the correct IP position, we introduced a "sharpening" process of the images.
For each pair of adjacent pixels, a new pixel was added halfway between them, and its value was linearly interpolated between the values of the two neighbor pixels. This process allowed us to preserve the hexagonal symmetry of the pixel matrix, while artificially increasing the sharpness of the images. In Fig.~\ref{fig:comp} an example of a track image before and after the sharpening process is shown. A detailed description of the process is reported in Appendix~\ref{app:a}.

Before giving the tracks as input to the network, images were reshaped in a 72x72 pixels frame. This size was chosen to reduce the training time as much as possible (the larger the images are, the larger the number of parameters in the network architecture, and the longer the time requested to process them) without cropping a significant number of tracks at high energies. We evaluated the learning ability of the network with a validation data set, and we chose to train the network for 60 epochs, with the OHEM process having been introduced in the 30th epoch. 

\subsection{Impact point reconstruction results}
\label{sec:ip_res}

After the training process, we tested our network with an independent data set, and we compared the results both with a network trained with unsharpened images and with the standard moment analysis. In order to lower the number of cropped tracks even further and to remove any possible bias introduced by the network, we first rotated each image by 120° and 240°, we then predicted the impact point for each of these three images, and finally we rotated the predictions back. These rotations allowed us to maintain unaltered the original hexagonal symmetry of the image. The reconstructed IP is the mean value of the three predictions made by the network.

\begin{figure}[ht]
    \centering
    \hspace*{-0.5cm}
    \includegraphics[width=10.2cm, height=10.2cm]{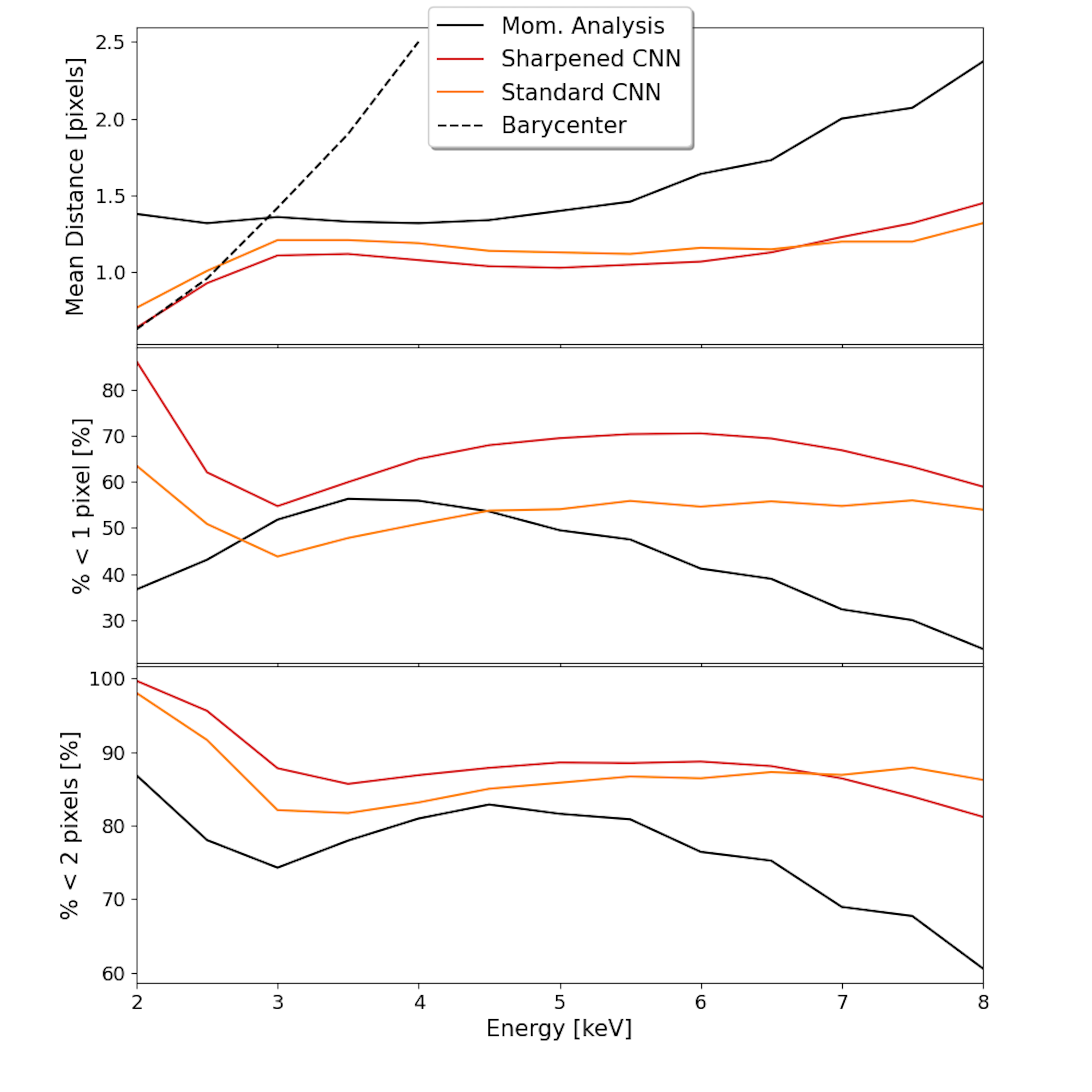}
    \caption{Comparison of the IP position reconstruction between our sharpened-image CNN (red line), an unsharpened-image CNN (orange line), and the moment analysis (black line). In the top panel, the mean distance between the true IP position and the predicted one is reported. The dashed black line represents the distance between the true IP and the barycenter of the track. In the middle and bottom panel we reported the percentage of events for which the distance between the true and the predicted IP position is lower than one and two pixels, respectively.}
    \label{fig:IP_comparison}
\end{figure}

To evaluate the performance of the models in the reconstruction of the IP position, we used three different figures of merit. We show the results as a function of energy in Fig.~\ref{fig:IP_comparison}. The mean distance between the predicted IPs and the true ones is reported in the top panel; the percentage of events for which the distance between the true IP position and the predicted one is smaller than one pixel (middle panel, \textit{$\% < 1$ pixel}) and two pixels (bottom panel, \textit{$\% < 2$ pixel}) are reported. The unit "pixel" we report in these figures of merit indicates the standard pixel dimension, not the sharpened one, for both the moment analysis and the two CNNs.

Firstly, the sharpening process allowed the CNN to reach a higher precision in the prediction of the IP location, as the mean distance for the sharpened CNN is lower than the one for the standard CNN in almost the entire energy range, while the percentage of events for which the distance is lower than one or two pixels is higher. For example, at 3 keV (6 keV) the \textit{$\% < 1$ pixel} is up by $\sim 11\%$ ($16 \%$) and the \textit{$\% < 2$ pixel} is up by $\sim 6\%$ ($2 \%$). It should be noted that the standard CNN overcomes the performance of the sharpened one at very high energy, and this is probably due to a higher number of very long tracks that are cropped with the sharpening process. This could be solved by extending the frame dimension, but it would result in slowing down the algorithm. As IXPE's effective area is very low at very high energies, we have given priority to speeding up the algorithm. However, if this needs to be used in other similar applications, this aspect might need further care. 

Moreover, the performance of the sharpened CNN is significantly better compared to the moment analysis for the entire energy range and according to all three figures of merit. For example, at 3 keV (6 keV) the \textit{$\% < 1$ pixel} is up by $\sim 3\%$ ($29 \%$) and the \textit{$\% < 2$ pixel} is up by $\sim 13\%$ ($12 \%$). It is interesting to notice how the \textit{$\% < 2$ pixel} for the CNN is consistently higher than $80\%$, showing how the network is also accurate in the identification of the IP position when the tracks become longer. This feature is key in order to reduce the radial modulation induced by the polarization leakage effect. 

In the top panel, the mean distance between the barycenter of the track and the true IP position is reported too. In the very low energy tracks, the true IP is very close to the barycenter, while the predictions of the moment analysis are likely to determine the IP position in a peripheral area of the track. Therefore, in the standard moment analysis, the prediction of the IP position could be manually substituted by the barycenter position to improve the precision at low energies. However, when handling  data without MC information, it is not trivial to select the energy threshold from where the barycenter should be employed as the predicted IP, and it is not applied in the standard analysis. The CNN, on the other hand, automatically follows the trend of the barycenter for low energies.

We conducted an analysis of the three simulated unpolarized point sources to determine if the improved detection of the IP could result in a reduction of the point spread function (PSF) of the instrument. By considering only the contributions to the PSF from the GPD, we found that the new CNN-predicted IP resulted in a half-power diameter (HPD), that is to say the diameter within which half of the collected X-rays are enclosed, that was approximately 30\% lower than the standard method \citep{IXPE_calibration}. However, the dominant contributors to the total PSF of IXPE are the mirror modules, and HPD is at least approximately three times higher than the GPD one. If we take all the contributors  into account, the overall improvement in the HPD employing the CNN-predicted IP was limited to only 1-2\%. \\

\section{Polarization results and discussion}
\label{sec:5}
\begin{figure}[ht]
    \centering
    \hspace*{-0.5cm}
    \includegraphics[width=10.2cm, height=7cm]{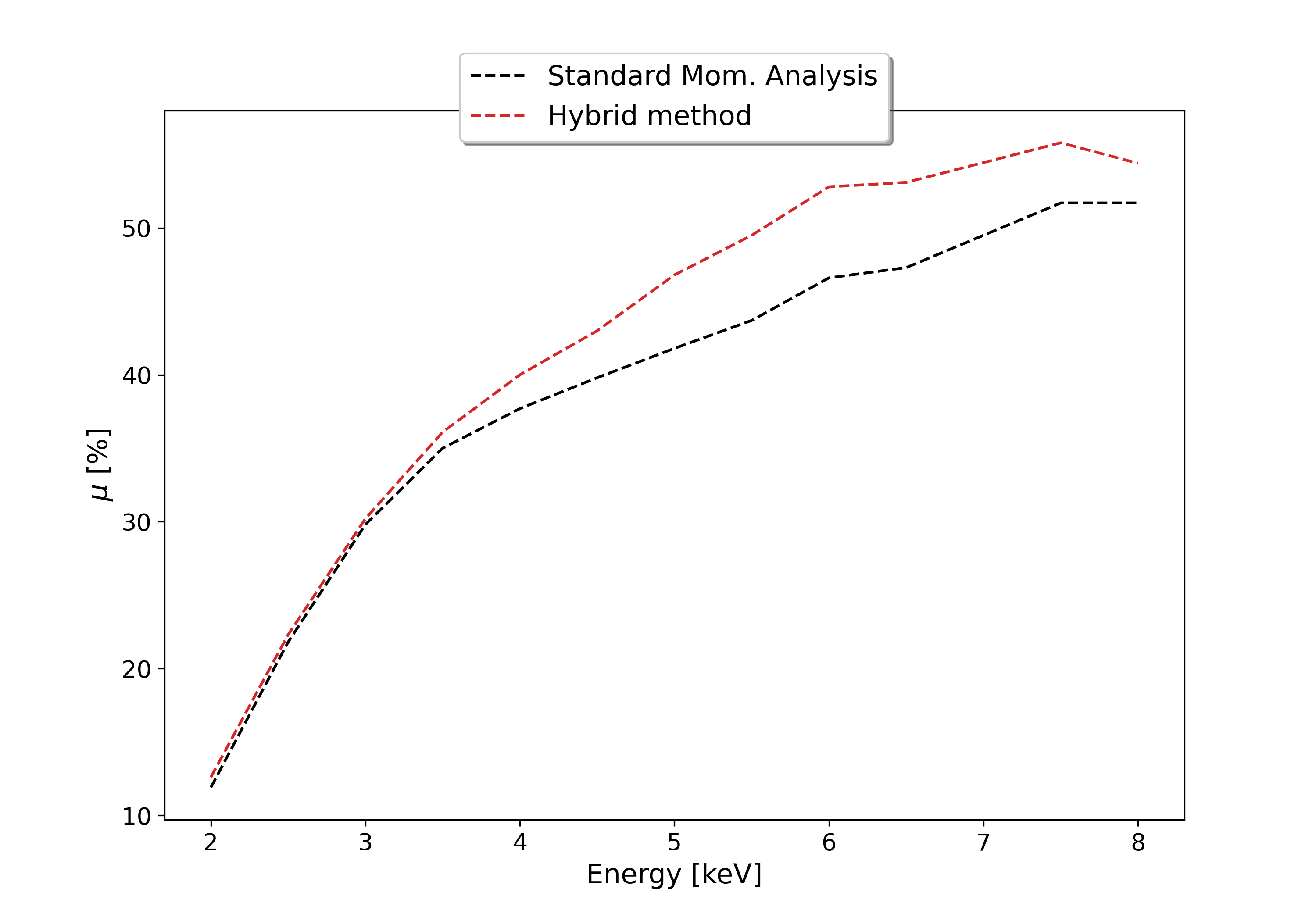}
    \caption{Modulation factor as a function of energy for the standard moment analysis (black) and for our hybrid algorithm (red).}
    \label{fig:ip_mom_hyb}
\end{figure}

The CNN-predicted IP was then passed to the moment analysis and introduced in Eq.~\ref{eq:mom_analysis} to predict the polarization of our testing samples and to evaluate the modulation factor as a function of energy. We report the results in Fig.~\ref{fig:ip_mom_hyb}.

Despite the substantially improved precision in the IP reconstruction at low energies ($\rm E < 3.5$ keV), the enhancement in the modulation factor value is marginal, around 1\%. At higher energies the improvement is more significant, up to $\sim$ 6\% at 6 keV. 

\begin{figure}[htb]
    \centering
    \includegraphics[width=7cm, height=6cm]{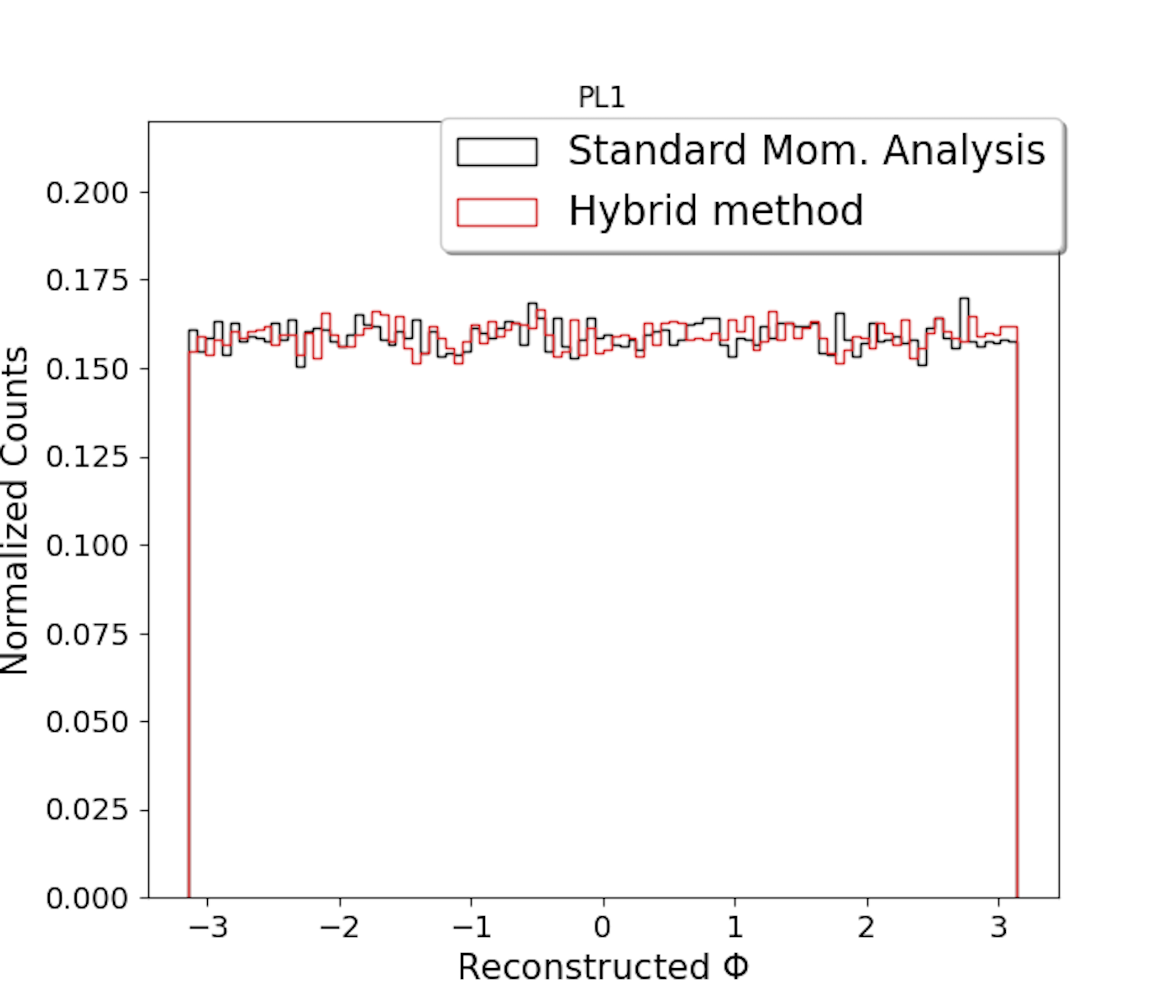}
    \includegraphics[width=7cm, height=6cm]{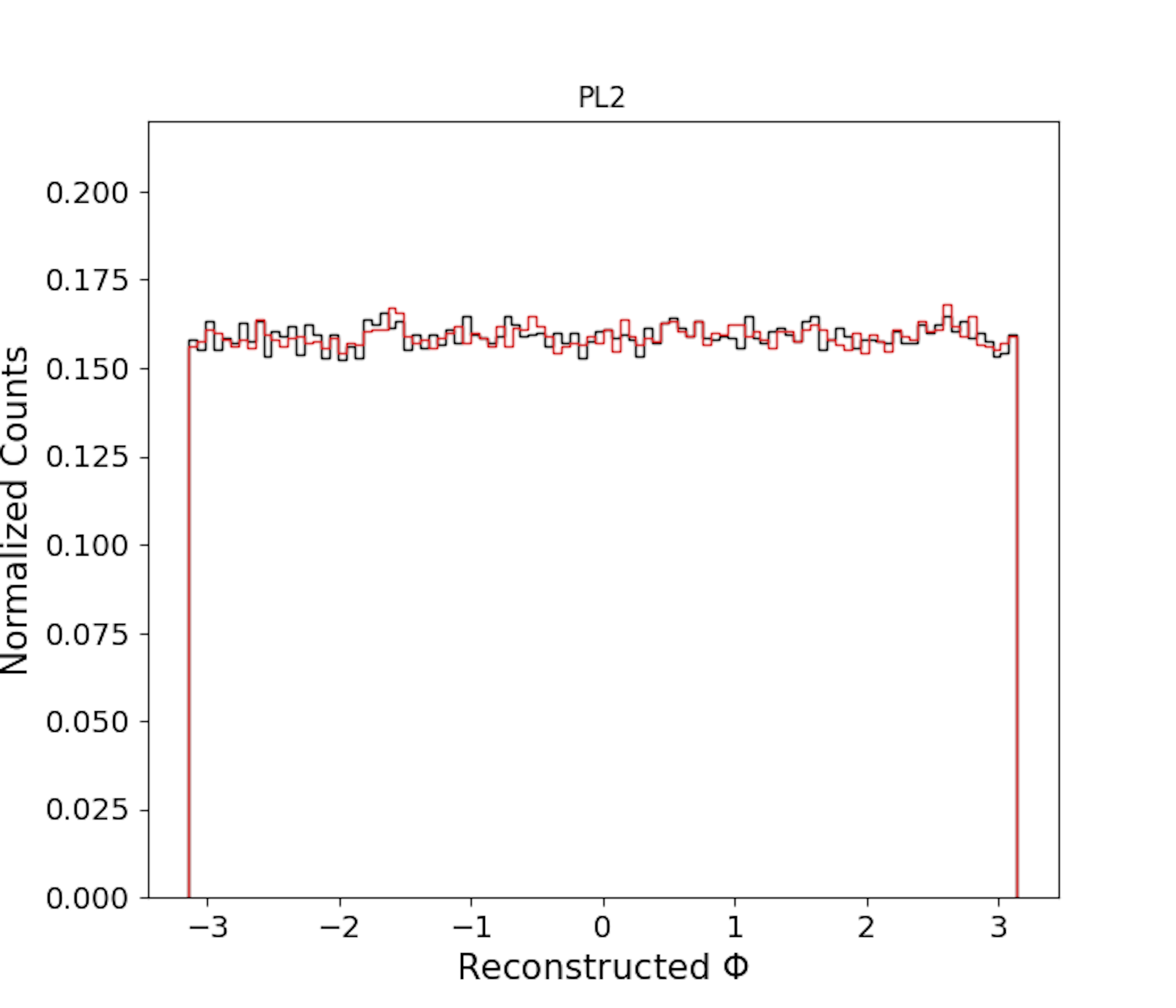}
    \includegraphics[width=7cm, height=6cm]{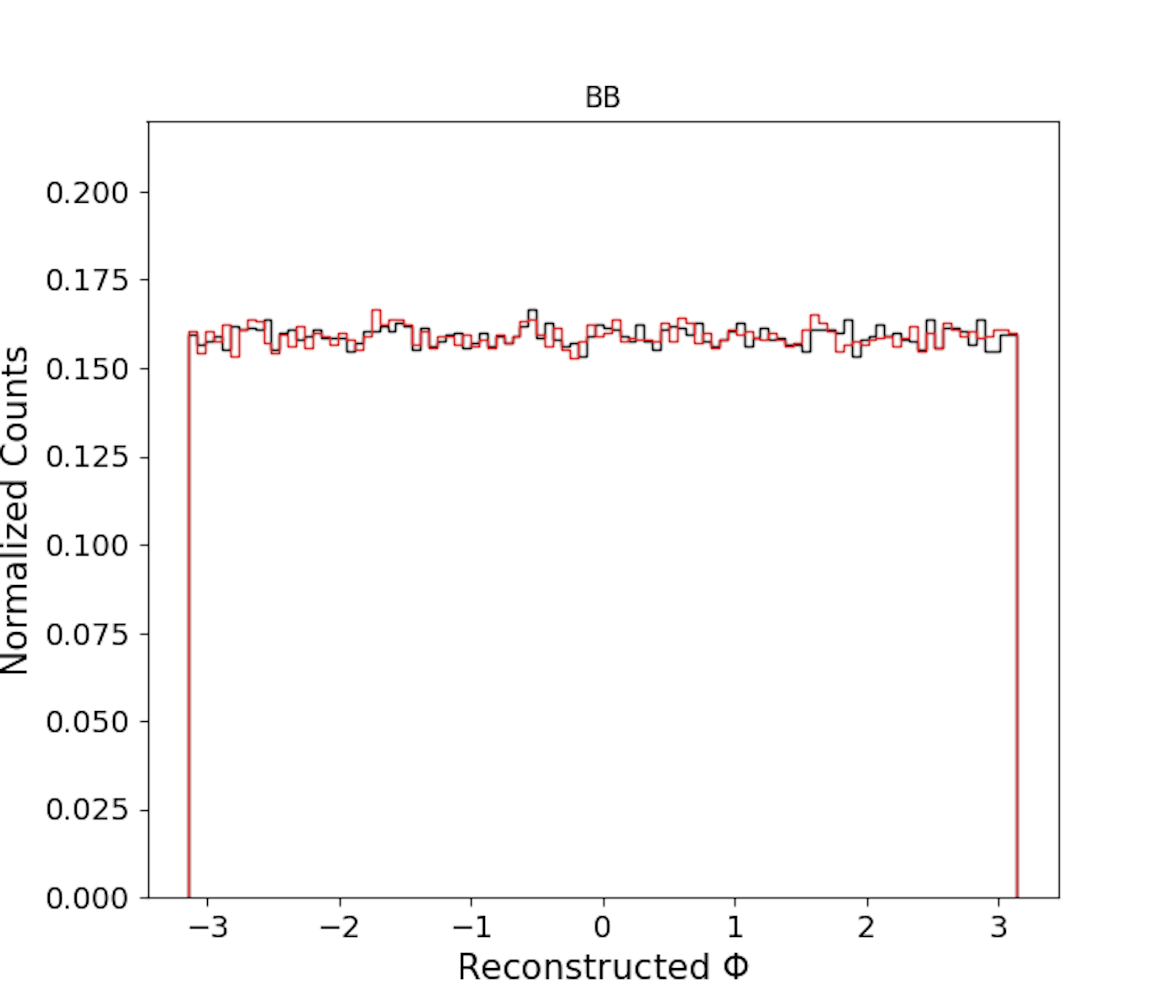}
    \caption{Histograms of the reconstructed emission angle for the standard moment analysis (black) and for the hybrid method (red), for the three unpolarized point sources. Modulation factor values are reported in Table~\ref{tab:1}.
    }
    \label{fig:hist_phi}
\end{figure}

We verified that the response to an unpolarized sample carries no residual modulation, by processing the three data sets simulating the unpolarized point sources (the two power-law spectra, PL1 and PL2, and the blackbody spectrum, BB). We calculated the residual modulation without aligning the predicted emission angle to the radial direction with the standard moment analysis and with the hybrid method. As expected, both algorithms found no residual modulation for all three sources, as reported in Fig.~\ref{fig:hist_phi} and in Table~\ref{tab:1} .

\begin{table}[ht]
\centering
\caption{Summary of the residual modulation for the unpolarized point sources with no radial direction projection. The results show no significant residual polarization, as all the modulation values are compatible with zero within $1\sigma$.}
\begin{adjustbox}{width=0.39\textwidth}
\begin{tabular}{ccc}
\hline
\hline
\noalign{\smallskip}
\textbf{Spectral Model} & \multicolumn{2}{c}{\textbf{Residual Linear Polarization} [\%]} \\
\hline
\noalign{\smallskip}
 & Mom. Analysis & Hybrid \\
 \noalign{\smallskip}
Power Law, index -1.7 (PL1) & $0.19 \pm 0.30$ & $0.15 \pm 0.30$\\
\noalign{\smallskip}
Power Law, index -0.7 (PL2) & $0.22 \pm 0.30$ & $0.17 \pm 0.30$\\
\noalign{\smallskip}
Blackbody, kT 3keV (BB) & $0.13 \pm 0.30$ & $0.18 \pm 0.30$ \\
\hline
\end{tabular}
\end{adjustbox}
\label{tab:1}
\end{table}

\begin{figure}[ht]
    \centering
    \hspace*{-0.6cm}
    \includegraphics[width=10.2cm, height=7cm]{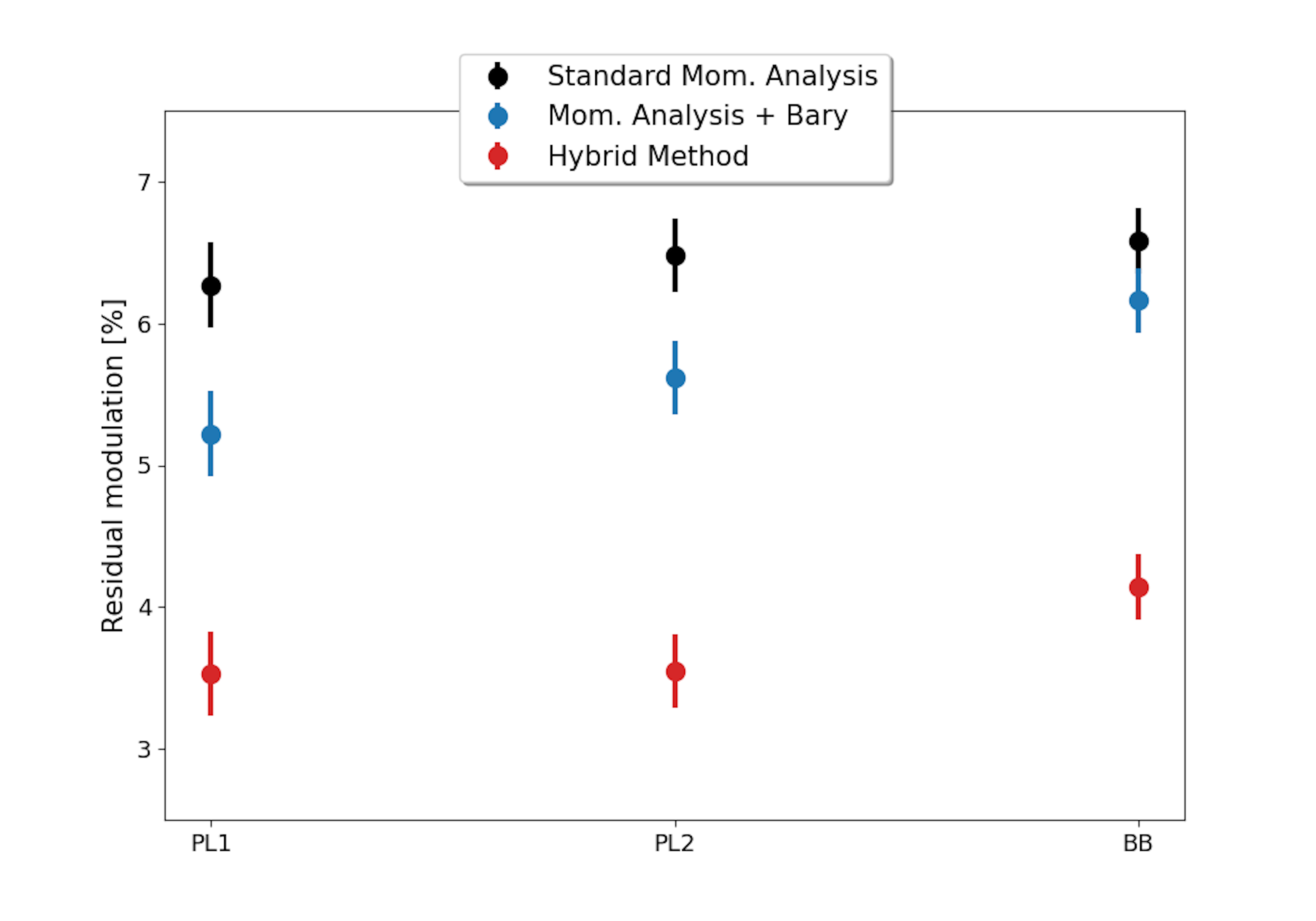}
    \caption{Residual radial modulation for the unpolarized point sources for three different methods: the standard moment analysis in black, the moment analysis that employs the barycenter as an IP prediction in blue, and our hybrid method in red.}
    \label{fig:pol_leak}
\end{figure}

We then analyzed the events of the unpolarized sources aligning the predicted emission angle to the radial direction to investigate the polarization leakage. We calculated the radial residual modulation for the standard moment analysis, for the same analysis but employing the barycenter as the best estimation of IP at low energies\footnote{As reported in the top panel of Fig.~\ref{fig:IP_comparison}, the barycenter is closer to the true IP with respect to the prediction of the moment analysis for energies lower than 3 keV.}, and for our hybrid method. The results are reported in Fig.~\ref{fig:pol_leak} and in Table~\ref{tab:2}.

\begin{table}[ht]
\centering
\caption{Summary of the residual radial modulation of the three unpolarized point sources.}
\begin{adjustbox}{width=0.49\textwidth}
\begin{tabular}{cccc}   
\hline
\hline
\noalign{\smallskip}
\textbf{Spectral Model} & \multicolumn{3}{c}{\textbf{Residual radial modulation} [\%]} \\
\hline
\noalign{\smallskip}
 & Mom. Analysis & Mom. Analysis + bary & Hybrid \\
 \noalign{\smallskip}
Power Law, index -1.7 (PL1)& $6.27 \pm 0.31$ &   $5.22 \pm 0.31$ &   $3.53 \pm 0.31$  \\ 
\noalign{\smallskip}            
Power Law, index -0.7 (PL2) & $6.48 \pm 0.26$ &   $5.62 \pm 0.26$ &   $3.55 \pm 0.26$ \\
\noalign{\smallskip}            
Blackbody, kT 3keV (BB) & $6.58 \pm 0.23$ &   $6.16 \pm 0.23$ &   $4.14 \pm 0.23$ \\
\hline
\end{tabular}
\end{adjustbox}
\label{tab:2}
\end{table}

As already mentioned in Sec.~\ref{sec::key2}, with the standard moment analysis, we measured residual modulations that are not compatible with zero, up to $6.58 \pm 0.23 \%$, even if the simulated point sources are unpolarized. We notice that employing the barycenter as the estimation of IP at low energies slightly mitigates this effect. 

With our hybrid method, we can reduce the residual modulation, by a factor up to approximately two with respect to the moment analysis, but that depends on the spectrum of the source considered. The effect is not completely eliminated, but the residual modulation is significantly reduced. One possible way to correct the polarization leakage effect involves an analytical PSF modeling \citep{leakage}, which, however, causes some residual errors. With our method, if the effect for the source is critical, corrections are still needed, but residuals would be lower.

\section{Conclusions}
\label{sec:6}

We built a new hybrid ML-analytic model for the reconstruction of the PE emission angle in a GPD with hexagonal pixels. Our results show a promising improvement in the performance compared to the state-of-the-art analytic algorithm. 

We implemented hexagonal convolutional layers as a C++ extension for PyTorch, which speeds up training and inference compared to pure Python/PyTorch implementation. We introduced a sharpening process of the images, resulting in a significant improvement in the reconstruction accuracy of the PE impact point by the CNN. We then modified the standard moment analysis by using this CNN-predicted impact point in Eq.~\ref{eq:mom_analysis}. Thanks to this change, our reconstruction algorithm performs better than the basic analytic reconstruction in recovering the modulation factor of 100\% polarized monochromatic beams at all tested energies (from $\sim$1\% at low energies to $\sim$6\% at higher energies), while it correctly predicts null linear polarization for the unpolarized sources. Moreover, employing the CNN-predicted IP in the standard moment analysis mitigates the polarization leakage effect up to a factor approximately two compared to the standard moment analysis. 

The results we have reported in this work do not consider any kind of weighting process of the events. As already shown in other works \citep{weights,ROMANI2021}, the performance could be further improved by applying an event-quality selection to remove or weight those events that convert outside the sensitive area of the detector. This topic, as well as the validation of the algorithm with real X-ray calibration data sets are currently being investigated and will be the subject of future study.

\begin{acknowledgements}
Portions of this research were conducted with high performance computing resources provided by Louisiana State University (http://www.hpc.lsu.edu).
We acknowledge Federica Legger, Sara Vallero, and the INFN Computing Center of Turin for providing support and computational resources, as well as the HPC4AI Laboratory of the University of Torino.
\end{acknowledgements}

\bibliographystyle{aa}
\bibliography{main}


\begin{appendix}

\section{Moment analysis, a step-by-step description}
\label{app:b}

As already mentioned in Sec. \ref{sec:ma}, the moment analysis aims to reconstruct analytically the track parameters exploiting its morphological properties. Here we provide the main steps of the analysis, while a more detailed description is reported in \cite{AnalyticRecon}.

Firstly, the barycenter of the charge distribution was calculated (blue dot in Fig. \ref{fig:rec_example}) as follows:
\begin{equation}
    x_{\rm{b}} = \frac{\sum_{\rm{i}} q_{\rm{i}} x_{\rm{i}}}{\sum_{\rm{i}} q_{\rm{i}}} \-\ \-\ \-\
    y_{\rm{b}} = \frac{\sum_{\rm{i}} q_{\rm{i}} y_{\rm{i}}}{\sum_{\rm{i}} q_{\rm{i}}}
,\end{equation}
\noindent
where $q_i$ is the charge collected in each $i_{th}$ pixel, and $(x_i, y_i)$ is the position of the pixel center on the readout plane. Defining $\phi$ as the angle respect to the $x$ axis, the second moment of the charge distribution $M_2(\phi)$ referred to the barycenter and to the direction defined by $\phi$ was then calculated:

\begin{equation}
    M_2(\phi)=\frac{\sum_{\rm{i}} q_{\rm{i}} [(x_{\rm{i}} - x_{\rm{b}}) \cos(\phi) + (y_{\rm{i}} - y_{\rm{b}}) \sin(\phi)]^2}{\sum_{\rm{i}} q_{\rm{i}}}~.
\end{equation}

The maximum and minimum values of $\rm M_2(\phi)$ correspond to the direction of the maximum and minimum elongation of the track, and they were obtained by imposing $\rm \frac{dM_2}{d\phi}=0$ (the dashed blue line in Fig. \ref{fig:rec_example} corresponds to the maximum elongation for the track). The third moment of the charge distribution
\begin{equation}
    M_3(\phi)=\frac{\sum_{\rm{i}} q_{\rm{i}} [(x_{\rm{i}} - x_{\rm{b}}) \cos(\phi) + (y_{\rm{i}} - y_{\rm{b}}) \sin(\phi)]^3}{\sum_{\rm{i}} q_{\rm{i}}}
    \label{eqn:mom_analysis}
\end{equation}
\noindent
allowed for the identification of the initial part of the track, as the PE lost more and more energy as it traveled through the gas, eventually forming the so-called "Bragg peak" when it was reabsorbed. Once the initial part of the track was selected, its barycenter was calculated. This new point was used to evaluate weights for each pixel in the whole track, as
\begin{equation}
    w_{\rm{i}} = \rm{e}^{-\frac{d_{\rm{b,i}}}{d_{\rm{s}}}}
,\end{equation}
where $d_{\rm{b,i}}$ is the distance between each track pixel and the position of the barycenter of the initial part of the track, and $d_{\rm{s}}$ is a scale parameter. The impact point was then defined as follows (green dot in Fig. \ref{fig:rec_example}):
\begin{equation}
    x_{\rm{IP}} = \frac{\sum_{\rm{i}} w_{\rm{i}} x_{\rm{i}}}{\sum_{\rm{i}} w_{\rm{i}}} \-\ \-\ \-\
    y_{\rm{IP}} = \frac{\sum_{\rm{i}} w_{\rm{i}} y_{\rm{i}}}{\sum_{\rm{i}} w_{\rm{i}}}~.
\end{equation}

The second moment of the charge distribution was again calculated, this time with respect to the location of the impact point $(x_{\rm{IP}}, y_{\rm{IP}})$ and with the weighted pixels,
\begin{equation}
    M_2'(\phi)=\frac{\sum_{\rm{i}} w_{\rm{i}} [(x_{\rm{i}} - x_{\rm{IP}}) \cos(\phi) + (y_{\rm{i}} - y_{\rm{IP}}) \sin(\phi)]^2}{\sum_{\rm{i}} w_{\rm{i}}}
    \label{eq:m2}
,\end{equation}
and the emission angle was obtained by evaluating the angle $\phi$ which maximizes $M_2'(\phi)$ (dashed green line in Fig. \ref{fig:rec_example}).
\\\\

\section{Artificial sharpening: Detailed description and considerations}
\label{app:a}

\begin{figure}[htb]
    \centering
    \includegraphics[width=9cm, height=9cm]{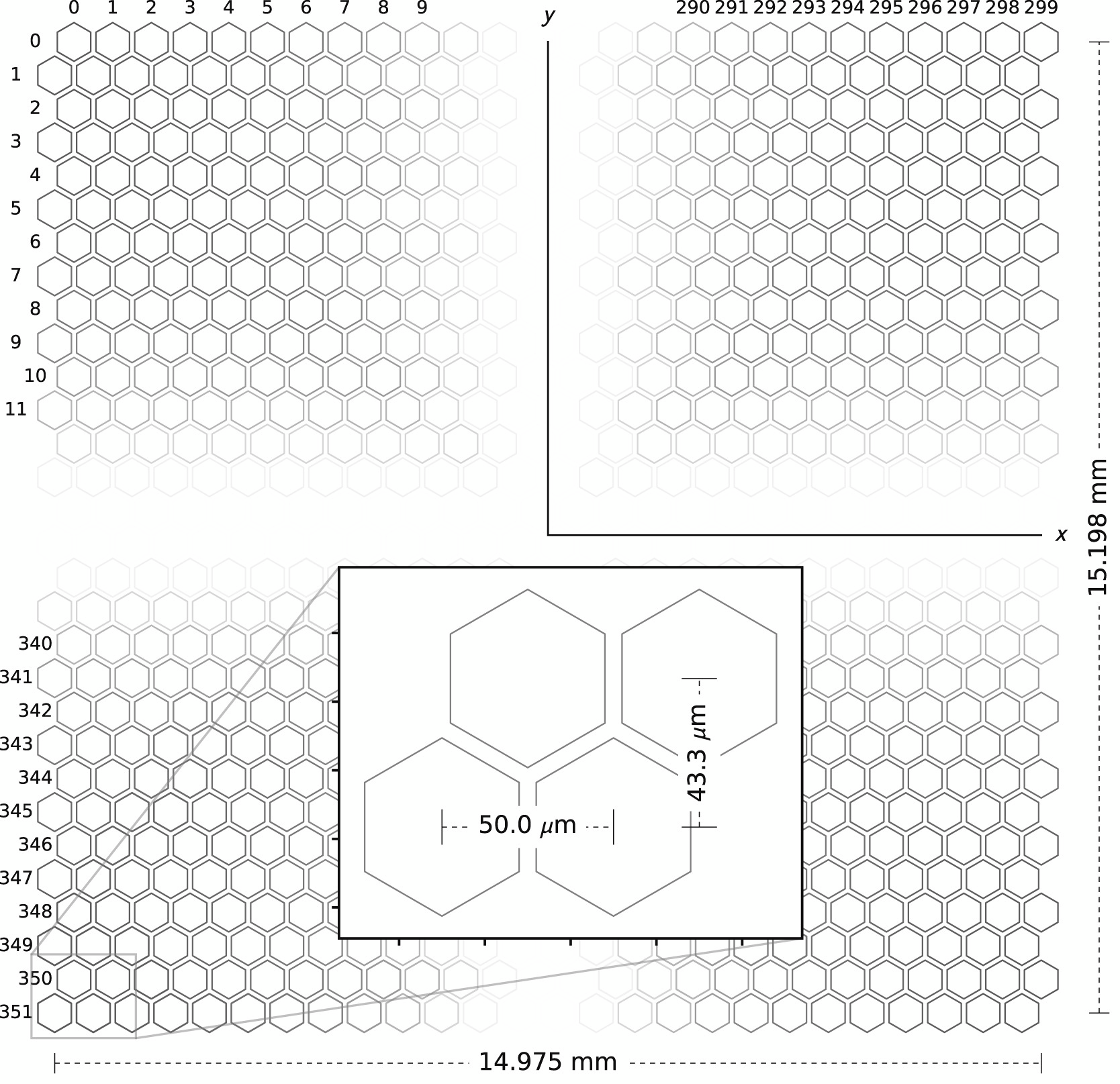}
    \caption{Scheme of the readout plane of the GPD. The $\rm (x, y)$ coordinate system has its origin in the center of the GPD, while the numbers on the borders of the pixel matrix refer to the $\rm (i, j)$ coordinate system. The horizontal and vertical distance between the center of the pixels is reported too. Image credits:~\cite{Coordinate}}
    \label{fig:asic_coord}
\end{figure}

The moment analysis requires the information on the impact point position to be in millimeters, while the output of our network is in the number of pixels. It is important then to build a map from one coordinate system to the other, and to modify it accordingly when sharpening the images.
The two coordinate systems are shown in Fig.~\ref{fig:asic_coord}~\citep{Coordinate}.

Knowing the number of rows and columns ($N_{\rm row}, N_{\rm col}$), and the horizontal and vertical distance between the centers of the hexagonal pixels ($p_{\rm row}, p_{\rm col}$), we can build a map from the position as a pixel number $(i, j)$ to the position in millimeters $(x, y)$:

\begin{equation}
    x = [i-\frac{1}{2}(N_{\rm{col}}-\frac{3}{2}+j \rm{mod} 2)] \cdot p_{\rm{col}}
\end{equation}
\begin{equation}
    y = [\frac{1}{2}(N_{\rm row}-1)-j] \cdot p_{\rm row}
.\end{equation}

When sharpening the images, a possible solution to preserve the spatial information of the track is to consider different GPD parameters and a change in the coordinate system.
Specifically: 
\begin{equation}\begin{split}
    N_{\rm col} \longrightarrow N'_{\rm col} = 2N_{\rm col} \-\ \-\ \-\ \-\  N_{\rm row} \longrightarrow N'_{\rm row} = 2N_{\rm row} \\
    p_{\rm col} \longrightarrow p'_{\rm col} = \frac{1}{2}p_{\rm col} \-\ \-\ \-\ \-\ p_{\rm row} \longrightarrow p'_{\rm row} = \frac{1}{2}p_{\rm row} \\
\end{split}\end{equation}
\begin{equation}
    i \longrightarrow i' = 2i - 1 \-\ \-\ \-\ \-\  j \longrightarrow j' = 2j
.\end{equation}

The $(x, y)$ position on the readout plane with respect to the new coordinates is the following:

\begin{equation}
    x = [i'-\frac{1}{2}(N'_{\rm col}-1+j' \rm mod2)] \cdot p'_{\rm col}
\end{equation}
\begin{equation}
    y = [\frac{1}{2}(N'_{\rm row}-2)-j'] \cdot p'_{\rm row}
.\end{equation}

Another factor that needs to be taken into account is that the image is passed to the network as a square image, and only is the CNN first convolutional layer interpreted as hexagonal. As we aim to achieve a high precision in the determination of the IP position, we need to correctly locate the IP on the square grid before giving the image as input to the network.

\begin{figure}[ht]
    \centering
    \includegraphics[width=9cm, height=6.7cm]{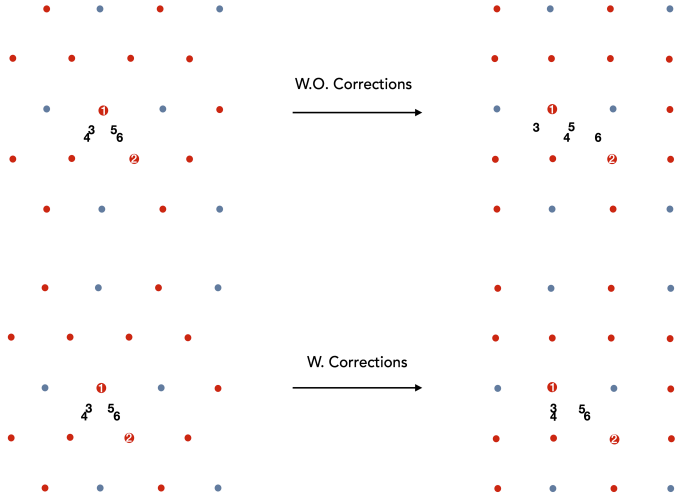}
    \caption{Scheme of an image pixel structure before and after the conversion from a hexagonal grid to a square grid. Each point represents a pixel center (the standard image pixels are reported in blue, and the ones added with the sharpening process are reported in red), while each number is an impact point position. In the top panel, the position of the IPs on the square grid without corrections is reported, while in the bottom panel the position of the IPs on the square grid takes a fine-tuning correction into account.}
    \label{fig:ip_shift}
\end{figure}

We consider six impact points in the same portion of an image in Fig.~\ref{fig:ip_shift}, where each point represents the center of a pixel and each number is a hypothetical impact point position. In blue the pixels of the standard images are reported, while in red the ones added with the sharpening process are provided. A positional bias could occur for an IP whose j value is not an integer, that is the IP is not located on a horizontal axis defined by the pixel center positions (IPs 3, 4, 5, and 6 in the same figure). After the $(x, y) \rightarrow (i, j)$ conversion, if we do not correct the $(i, j)$ values, the six IPs will be located on the square grid as in the upper panel of Fig.~\ref{fig:ip_shift}. 
A correction of the IP positions proportional to their distance from the closest integer j value is needed to obtain a configuration as in the lower panel of Fig.~\ref{fig:ip_shift}, which is more consistent with the positions of the IP on the hexagonal grid.

\end{appendix}
\end{document}